\documentclass{PoS}
\usepackage[utf8]{inputenc}
\usepackage{amssymb}
\usepackage{amsmath}
\usepackage{amstext}
\usepackage{bbm} 
\usepackage{slashed} 
\usepackage{booktabs} 
\usepackage{csquotes} 
\usepackage{subcaption}
\usepackage{cancel}
\usepackage{tikz}
\usepackage{wrapfig}
\usetikzlibrary{decorations.pathreplacing}
\usetikzlibrary{decorations.markings}

\newcommand{\ii}{\mathrm{i}}
\newcommand{\e}{\mathrm{e}}
\newcommand{\tr}{\mathrm{tr}}

\title{$\mathcal{N}=1$ Supersymmetric SU(3) Gauge Theory with a Twist}

\ShortTitle{$\mathcal{N}=1$ Supersymmetric SU(3) Gauge Theory with a Twist}

\author{\speaker{Marc Steinhauser}\\
        Friedrich Schiller University Jena, 07743 Jena, Germany\\
        E-mail: \email{marc.steinhauser@uni-jena.de}}

\author{Andre Sternbeck\\
        Friedrich Schiller University Jena, 07743 Jena, Germany\\
        E-mail: \email{andre.sternbeck@uni-jena.de}}
        
\author{Bj\"orn Wellegehausen\\
        Friedrich Schiller University Jena, 07743 Jena, Germany\\
        E-mail: \email{bjoern.wellegehausen@uni-jena.de}}
        
\author{Andreas Wipf\\
        Friedrich Schiller University Jena, 07743 Jena, Germany\\
        E-mail: \email{wipf@tpi.uni-jena.de}}

\abstract{We investigate the pure gauge sector of Super-QCD, i.e. Super-Yang-Mills (SYM) theory, with focus on the bound states. To improve chiral symmetry as well as supersymmetry at finite lattice spacing, we use a deformed SYM lattice action. It contains a twist term, similar to the lattice formulation of twisted mass QCD. We present the status of our theoretical and numerical investigation.
}

\FullConference{37th International Symposium on Lattice Field Theory - Lattice2019\\
		16-22 June 2019\\
		Wuhan, China}

\begin{document}

\section{Introduction}
Although the standard model of particle physics is very successful in predicting a broad range of observables, there are open questions which it can not explain. A popular extension of the standard model includes supersymmetry. A supersymmetric extension could solve, for example, the hierarchy problem of the Higgs mass and the lightest supersymmetric particle (LSP) provides a candidate for dark matter \cite{Witten:1981nf,Dimopoulos:1981zb,Ellis:1983ew}.

For a better understanding of the non-perturbative low-energy regime of this model, we investigate the gauge part which just contains the gluons and their superpartners, the gluinos. This supersymmetric extension of Yang-Mills theory (YM) is called $\mathcal{N}\!=\!1$ Super-Yang-Mills theory (SYM) and its on-shell Lagrange density
\begin{equation}
	\mathcal{L}_\text{SYM} = \tr \left( -\frac{1}{4}F_{\mu\nu} F^{\mu\nu} + \frac{\ii}{2} \bar{\lambda} \slashed{D} \lambda - \frac{m_\text{g}}{2} \bar{\lambda}\lambda \right)
	\label{eq:Langrangian}
\end{equation}
looks similar to QCD but with a single flavor. Tough, there are more differences: The fermionic gluino is described by a Majorana field $\lambda(x)$ which transforms in the adjoint representation like the gauge potential $A_\mu(x)$. Supersymmetry dictates these properties to guarantee the same number of bosonic and fermionic degrees of freedom. On the lattice, we use Wilson fermions for $\lambda$. The mass term in eq.~\eqref{eq:Langrangian} breaks supersymmetry softly, but is used to fine-tune the action to vanishing gluino mass. The supersymmetry transformation
\begin{equation}
	\delta_\epsilon A_\mu = \ii \bar{\epsilon} \gamma_\mu \lambda,\qquad\delta_\epsilon\lambda = \ii\Sigma_{\mu\nu} F^{\mu\nu} \epsilon
	\label{eq:SusyTrafo}
\end{equation}
transforms the fermionic and bosonic fields into each other. In the previous equation, $\epsilon$ is an infinitesimal Majorana spinor and $\Sigma_{\mu\nu} \equiv \frac{\ii}{4} \left[ \gamma_\mu, \gamma_\nu \right]$.
\begin{figure}
	\begin{minipage}{0.45\textwidth}
	The low-lying bound states of the SYM model were first predicted by Veneziano and Yankielowicz in the framework of effective field theory and based on the symmetries and anomalies of the theory  \cite{Veneziano8206}. This multiplet contains two mesonic states $\text{a-}\eta^\prime$ \& $\text{a-}f_0$, named after their QCD-counterparts, and a gluino-glueball $gg$. When supersymmetry is unbroken ($m_\text{g}=0$ in eq.~\eqref{eq:Langrangian}) all states within a supermultiplet are mass-degenerated. If the gluino mass is switched on, the masses of the states are shifted, cf. Figure~\ref{fig:VYmultiplet}.\\
	\end{minipage}
	\hfill
	\begin{minipage}{0.47\textwidth}
		\centering
		\footnotesize
		\begin{tikzpicture}[>=latex,every text node part/.style={align=center},every node/.style={scale=1.2},scale=0.45]
		\draw[->,line width=0.6mm] (4,5) -- (4,12) node[above]{mass};
		\draw[line width=0.5mm] (5,7.75) -- (8,7.75);
		\draw[line width=0.3mm, dotted] (8,7.75) -- (10,8.75);
		\draw[line width=0.5mm] (10,8.75) -- (13,8.75) node[right]{$\text{a-}f_0$};
		\draw[line width=0.3mm, dotted] (8,7.75) -- (10,7.75);
		\draw[line width=0.5mm] (10,7.75) -- (13,7.75) node[right]{$~gg$};
		\draw[line width=0.3mm, dotted] (8,7.75) -- (10,6.75);
		\draw[line width=0.5mm] (10,6.75) -- (13,6.75) node[right]{$\text{a-}\eta^\prime$};
		\node at (6.5,11) {susy};
		\node at (11.5,11) {softly\\broken\\susy};
		\node at (6.5,5) {$m_\text{g}=0$};
		\node at (11.5,5) {$m_\text{g}\neq0$};
		\end{tikzpicture}
		\caption{Sketch of the Veneziano-Yankielowicz multiplet for unbroken ($m_\text{g}=0$) and softly broken ($m_\text{g}\neq0$) supersymmetry.}
		\label{fig:VYmultiplet}
	\end{minipage}
\end{figure}
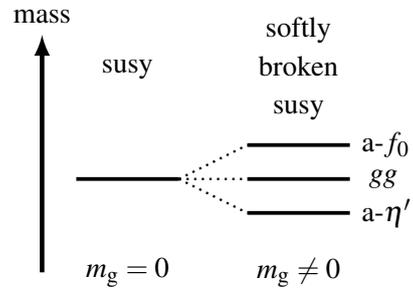\\
The massless SU(3) SYM action is invariant under $\lambda\mapsto\exp(\ii\alpha\gamma_5)\lambda$. This global chiral U$(1)_\text{A}$ symmetry is anomalous and only a $\mathbb{Z}_6$ symmetry remains,
\begin{equation}
	\lambda\mapsto\exp(\ii\alpha\gamma_5)\lambda~~\text{with}~~\alpha=2\pi n/6,~~n\in\{1,\ldots,6\}\,.
	\label{eq:ChiralSymmetry}
\end{equation}
Furthermore, a non-vanishing gluino condensate $\langle \bar{\lambda}\lambda \rangle \neq 0$ spontaneously breaks this down to a $\mathbb{Z}_2$ symmetry \cite{Shifman:1987ia}.

\section{Lattice formulation with a twist}
Curci and Veneziano proposed for the $\mathcal{N}\!=\!1$ SYM theory a lattice formulation with Wilson fermions \cite{Curci8612}.
At finite lattice spacing, supersymmetry is broken but a fine-tuning of the gluino mass assures its restoration in the continuum limit. In analogy to twisted-mass QCD we add a parity breaking-mass $\ii\mu\gamma_5\delta_{x,y}$ to the Dirac operator\footnote{Note the adjoint representation $\left[ \mathcal{V}_\mu(x)\right] _{ab}=2\,\tr\left[\mathcal{U}_\mu^\dagger(x)T_a\,\mathcal{U}_\mu(x) T_b \right]$ of the gauge field.}:
\begin{align}
	D_\text{W}^{\text{mtw}}(x,y)\!&=\!(4+m+\ii\mu\gamma_5)\delta_{x,y} - \frac{1}{2}\!\sum_{\mu=\pm1}^{\pm4}\left( \mathbbm{1}-\gamma_\mu \right)\!\mathcal{V}_\mu(x)\,\delta_{x+\hat{\mu},y}\nonumber\\
	&=\!(4+\,~M~\,\e^{\ii\alpha\gamma_5}~)\delta_{x,y} - \frac{1}{2}\!\sum_{\mu=\pm1}^{\pm4}\left( \mathbbm{1}-\gamma_\mu \right)\!\mathcal{V}_\mu(x)\,\delta_{x+\hat{\mu},y}\,.
	\label{eq:DiracOperatorWithTwist}
\end{align}
Our initial motivation for this formulation was a feature of SYM. Particular directions of the $\mathbb{Z}_{6}$ symmetry (cf.~eq.~\eqref{eq:ChiralSymmetry}) are favored by the gluino condensate. The mass $m=M\,\cos(\alpha)$ breaks the chiral symmetry explicitly and generates a condensate $\sim\langle\bar{\lambda}\lambda\rangle$. Additionally $\mu=M\sin(\alpha)$ leads to a condensate $\sim\langle\bar{\lambda}\gamma_5\lambda\rangle$, connected to the former by a U(1)-symmetry. 

In contrast to QCD, where the twisted basis is rotated back to the physical basis for the observables, we keep the deformation of the $\mu$-term. This deformation of the lattice action vanishes in the chiral limit $m\rightarrow m_\text{crit},~\mu\rightarrow0$. The chiral point is approached along a specific \enquote{direction} in the $(m,\mu)$-plane. This may be beneficial for the continuum extrapolation because there are indications we are closer to chiral symmetry and supersymmetry at finite lattice spacing.

Independent lattice calculations of the $\mathcal{N}\!=\!1$ SYM theory with an untwisted lattice action are performed by the DESY-M\"unster collaboration. Their investigations cover the spectrum for gauge group SU(3) \cite{Ali:2019agk,Bergner:2019}, variational analysis of the SU(2) spectrum \cite{Ali:2019gzj,Scior:2019} and the chiral symmetry in thermal SU(2) $\mathcal{N}\!=\!1$ SYM theory using the gradient flow \cite{Bergner:2019dim,Lopez:2019}.

\section{Results}\label{ch:Results}

\subsection{Fermionic Expectation Values}
For any observable $O$ the expectation value\footnote{The fermionic measure does not contain $\mathcal{D}\bar{\lambda}$. The Majorana condition relates $\lambda=\lambda^\text{c}=\mathcal{C}\bar{\lambda}^\text{T}$ with $\bar{\lambda}$ through the charge conjugation.}
\begin{align*}
	\langle O\, \rangle &= \langle \langle O\, \rangle_\text{F} \rangle_\text{G}
	= \frac{1}{Z} \int \mathcal{DU}\, \e^{-S_\text{G}[\mathcal{U}]}\, \mathcal{D}\lambda\, \e^{-S_\text{F}[\lambda,\mathcal{U}]}\, O[\lambda,\mathcal{U}]
\end{align*}
contains a field integration over the gauge field $\mathcal{U}$ and the Majorana field $\lambda$. Performing a twist, the spinors are transformed $\lambda\mapsto\e^{\ii \alpha\gamma_5/2}\lambda$ and $\bar{\lambda}\mapsto\bar{\lambda}\e^{\ii \alpha\gamma_5/2}$, which can be combined to
\begin{align*}
	\begin{pmatrix}
	\bar{\lambda}\lambda\\ 
	\ii\bar{\lambda}\gamma_5\lambda
	\end{pmatrix}\!\!\!\!&~~\mapsto
	\begin{pmatrix}
	~~~\cos(\alpha)& \sin(\alpha) \\ 
	-\sin(\alpha)& \cos(\alpha)
	\end{pmatrix} 
	\begin{pmatrix}
	\bar{\lambda}\lambda\\ 
	\ii\bar{\lambda}\gamma_5\lambda
	\end{pmatrix}\,.
\end{align*}
We parametrize a generic mesonic interpolator by $O_{\text{gen}}(a,b)=a\,\bar{\lambda}_x\lambda_x+b\,\bar{\lambda}_x\gamma_5\lambda_x$. Then, mesonic interpolators of the Veneziano-Yankielowicz multiplet transform as
\begin{align*}
O_{\text{a-}f_0}(1,0)&=\bar{\lambda}_x\lambda_x ~~~\mapsto O_{\text{gen}}(\cos(\alpha),\ii\sin(\alpha))\\
O_{\text{a-}\eta^\prime}(0,1)&=\bar{\lambda}_x\gamma_5\lambda_x\mapsto O_{\text{gen}}(\ii\sin(\alpha),\cos(\alpha))\,.
\end{align*}
Consequently the mesonic correlators
\begin{align*}
\langle O_{\text{a-}f_0}(n) \bar{O}_{\text{a-}f_0}(m) \rangle_\text{F}(0^\circ) \, &= \langle O_{\text{a-}\eta^\prime}(n) \bar{O}_{\text{a-}\eta^\prime}(m) \rangle_\text{F}(90^\circ) \text{~~and}\\
	\langle O_{\text{a-}\eta^\prime}(n) \bar{O}_{\text{a-}\eta^\prime}(m) \rangle_\text{F}(0^\circ)&=\langle O_{\text{a-}f_0}(n) \bar{O}_{\text{a-}f_0}(m) \rangle_\text{F}(90^\circ)
\end{align*}
change their roles when the angle $\alpha=90^\circ$ is chosen. In between (at $\alpha=45^\circ$) we find
that the expectation values $\langle O_{\text{a-}\eta^\prime}(n) \bar{O}_{\text{a-}\eta^\prime}(m) \rangle_\text{F}(45^\circ)$ and $\langle O_{\text{a-}f_0}(n) \bar{O}_{\text{a-}f_0}(m) \rangle_\text{F}(45^\circ)$ are similar to each other.
In other words, the chiral part of our supermultiplet has the same mass. Furthermore, the chiral rotation with the angle $\alpha$ can be interpreted as a deformation of our Dirac operator. We present in section~\ref{ch:MesonicStates} numerical evidence that approaching the chiral point along the \enquote{direction} $\alpha=45^\circ$ is beneficial. Although the supersymmetry is broken at finite lattice spacing, the mesonic states of the multiplet are mass-degenerated within the measurement uncertainty.

\subsection{Eigenvalues}
To understand the consequences of the twist we analytically investigate the free Dirac operator and its eigenvalues. Here, we introduce for generality a second twist $\varphi$ to the Dirac operator\footnote{A general action of this form was presented in \cite{Immirzi:1982hj}. In \cite{bergner_low-dimensional_2008}, this idea lead to $\mathcal{O}(a^3)$ improvement in the two-dimensional lattice formulation of the Wess-Zumino model.}
\vspace*{-2mm}\begin{equation}
	D_\text{W}^\text{dtw}(x,y)\!=\!(4\,\e^{\ii\varphi\gamma_5}+M\,\e^{\ii\alpha\gamma_5})\delta_{x,y} - \frac{1}{2}\!\sum_{\mu=\pm1}^{\pm4}\left( \e^{\ii\varphi\gamma_5}-\gamma_\mu \right)\!\mathcal{V}_\mu(x)\,\delta_{x+\hat{\mu},y}
	\label{eq:dtw}
	\vspace*{-2mm}
\end{equation}
and quantify discretization errors in the lattice spacing $a$. Our results for several fermionic kernels $D$ are summarized in Table~\ref{tab:eigenvalues}. The unimproved Dirac-operator $D_1$ (as defined in Table~\ref{tab:eigenvalues}) has $\mathcal{O}(a)$ discretization effects, which can be avoided by a twist $\alpha=90^\circ$ in $D_2$ (i.e. fully twisted lattice QCD) or with a modified Wilson term in $D_3$. In the general case
of $D_4$ any combination \mbox{$\alpha-\varphi=90^\circ~(\text{mod}~180^\circ)$} leads as well to $\mathcal{O}(a)$ improvement. To summarize, improvement of the discretization effects is achieved if the mass term and Wilson term are orthogonal to each other.
\begin{table}[h]
	\caption{Analytic results for the eigenvalues $\lambda^\dagger\lambda$ of several fermionic kernels $D_i$. The results are expanded in the lattice spacing $a$ and we defined  $\kappa=-\frac{1}{3}\sum_\mu p_\mu^4+\frac{r^2}{2}\Big(\sum_\mu p_\mu^2\Big)^2$.}
	\centering
	\begin{tabular}{l|l}
	Fermionic Kernel & Eigenvalues $\lambda^\dagger\lambda$\\ \hline\\[-1.0em]
	$D_1=\gamma^\mu\partial_\mu + m - \frac{r}{2} \Delta$ & $p^2+m^2+amrp^2+\mathcal{O}(a^2)$ \\[-0.25em]
	$D_2=\gamma^\mu\partial_\mu + M\,\e^{\ii\alpha\gamma_5} - \frac{r}{2} \Delta$	& $p^2+M^2+\cancelto{0~\text{for}~\alpha=90^\circ}{aMrp^2\cos(\alpha)}+\mathcal{O}(a^2)$ \\[+0.25em]
	$D_3=\gamma^\mu\partial_\mu + m + \frac{\ii r}{2}\gamma_5\Delta$ & $p^2+m^2+\kappa a^2\,+\mathcal{O}(a^4)$ \\[-0.5em]
	$D_4=\gamma^\mu\partial_\mu + M\,\e^{\ii\alpha\gamma_5} + \frac{r}{2}\e^{\ii\varphi\gamma_5}\Delta$ & $p^2+M^2+\cancelto{0~\text{for}~\alpha-\varphi=90^\circ}{aMrp^2\cos(\alpha-\varphi)}+\kappa a^2+\mathcal{O}(\cancelto{\,a^4}{a^3})$
	\end{tabular}
	\label{tab:eigenvalues}
\end{table}

One angle in the description \eqref{eq:dtw} is redundant and can be transformed away. Thus we have to decide which angle is best suited for numerical investigations. In our opinion, the improvement of the mass-degeneracy in the supermultiplet at $\alpha=45^\circ$ together with a reduction of $\mathcal{O}(a)$ discretization effects by a factor $\cos(45^\circ)\approx0.7$ is more beneficial than the complete $\mathcal{O}(a)$ improvement at maximal twist $\alpha=90^\circ$.


\subsection{Connected Correlators}
To demonstrate the features of our twisted Dirac operator (eq.~\eqref{eq:DiracOperatorWithTwist}) we performed a parameter scan in the $(m,\mu)$-parameter space. Confinement prevents direct measurement of the gluino mass to find the chiral point, where in the continuum limit the chiral symmetry and the supersymmetry are simultaneously restored. Instead the pion\footnote{In the $\mathcal{N}\!=\!1$ SYM theory the pion is not part of the physical spectrum, but can defined in a partially-quenched framework and corresponds to the connected part of $\text{a-}\eta^\prime$,} mass $m_{\text{a-}\pi}^2\sim m_\text{g}$ is used for fine-tuning \cite{Veneziano8206}.
\begin{figure}[h!]
	\begin{minipage}[c]{0.33\textwidth}
		\includegraphics[height=4.0cm]{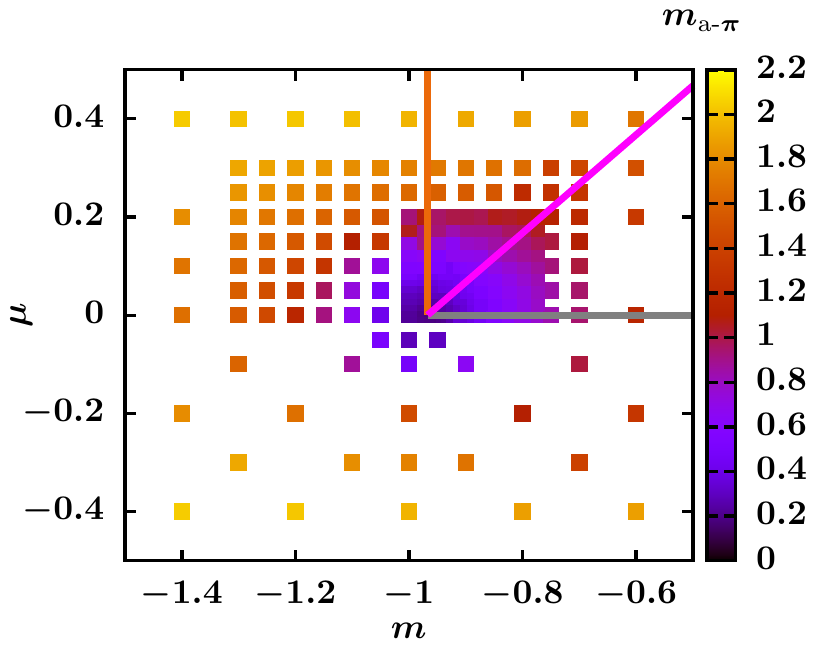}
	\end{minipage}
	\begin{minipage}[c]{0.33\textwidth}
		\includegraphics[height=4.0cm]{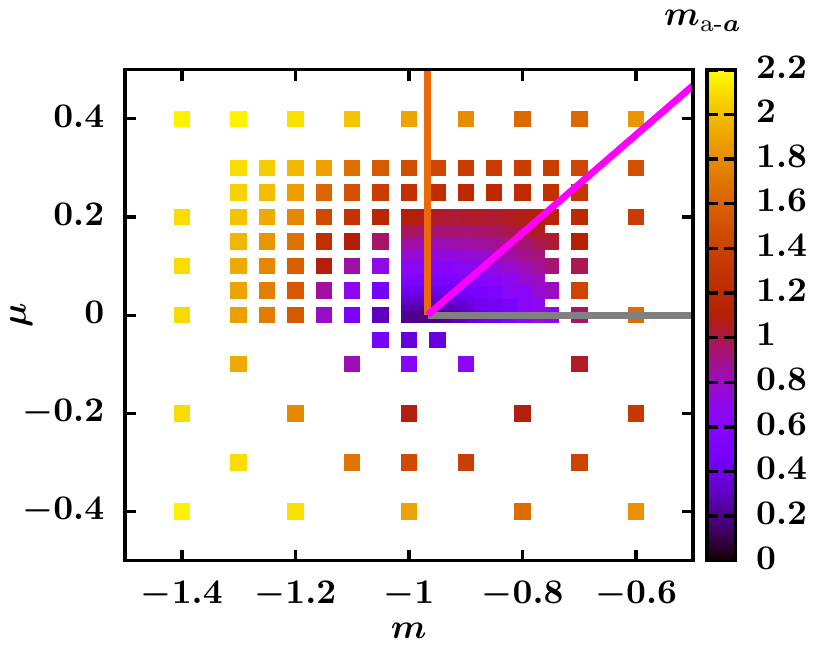}
	\end{minipage}
	\begin{minipage}[c]{0.33\textwidth}
		\includegraphics[height=4.0cm]{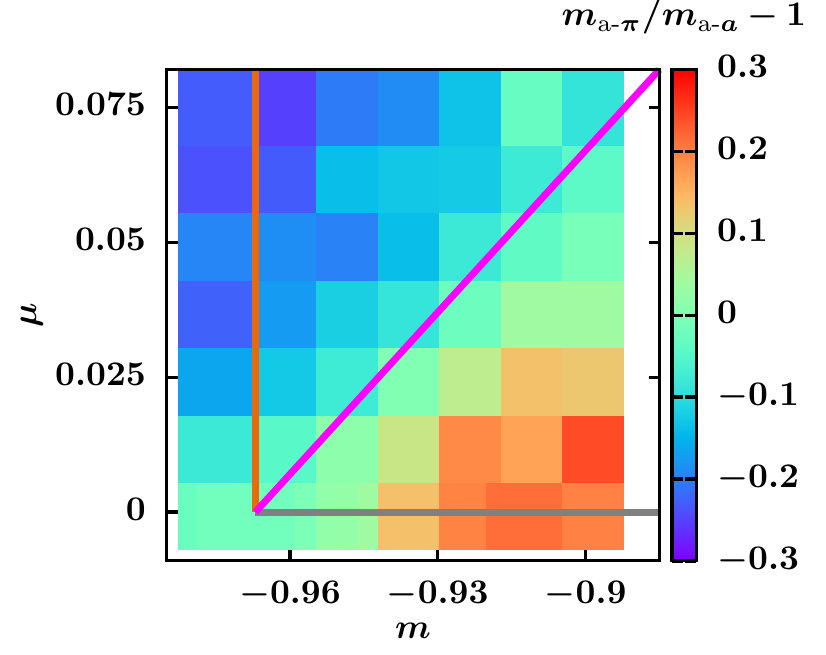}
	\end{minipage}
	\caption{Parameter scan in the $(m,\mu)$-plane on the $8^3\times 16$ lattice. The left plot shows the mass of the $\text{a-}\pi$ (connected part of the $\text{a-}\eta^\prime$), the middle plot depicts the mass of the $\text{a-}a$ (connected part of the $\text{a-}f_0$) and the right plot combine those masses as $m_{\text{a-}\pi}/m_{\text{a-}a}-1$. The colored lines in gray, magenta and orange are discussed in the text.}
	\label{fig:PiA}
\end{figure}
\begin{figure}[h!]
	\centering
	\includegraphics[width=0.7\textwidth]{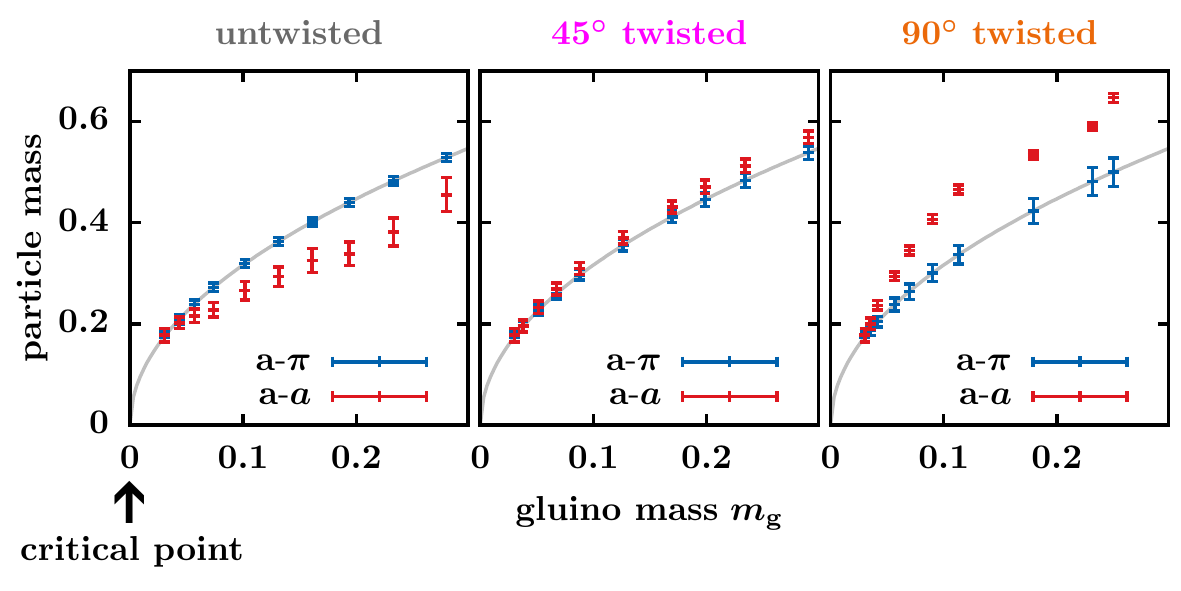}
	\caption{The same data as in Figure~\ref{fig:PiA}, but as cuts along the three directions depicted by the lines in gray, magenta and orange. The three plots corresponds to twist angles $\alpha\in\{0^\circ,45^\circ,90^\circ\}$ and show a clear difference in their mass hierarchies.}
	\label{fig:PiACuts}
\end{figure}

In Figure~\ref{fig:PiA} there are three directions towards the chiral limit, $m\rightarrow m_\text{crit}$.
Without a twist ($\mu=0$), one approaches $m\rightarrow m_\text{crit}$ along the gray line. Further interesting \enquote{directions} are the magenta line with $\alpha=45^\circ$ and the fully twisted orange line with $\alpha=90^\circ$. For the broad parameter scan we used the connected part of the $\text{a-}\eta^\prime$ called $\text{a-}\pi$ and the connected part of the $\text{a-}f_0$ called $\text{a-}a$ as observables approximating those states of the supermultiplet, which are expected to be mass-degenerated in the continuum theory. The right panel of Figure~\ref{fig:PiA} shows the subtracted ratio $m_{\text{a-}\pi}/m_{\text{a-}a} - 1 $ and reveals three different areas. The red/blue region indicates that the $\text{a-}\pi$ is heavier/lighter than the $\text{a-}a$. In between, in the green region, both states have comparable masses. In Figure~\ref{fig:PiACuts} the same data is depicted along the three \enquote{directions} $\alpha=\{0^\circ,45^\circ,90^\circ\}$ corresponding to the gray/magenta/orange line. This figure indicates, that for $\alpha=45^\circ$ improvement of the chiral symmetry and supersymmetry at finite lattice spacing is realized.

\subsection{Mesonic States}\label{ch:MesonicStates}


\begin{figure}
	\begin{minipage}{0.47\textwidth}
		\flushleft
		\includegraphics[width=\textwidth]{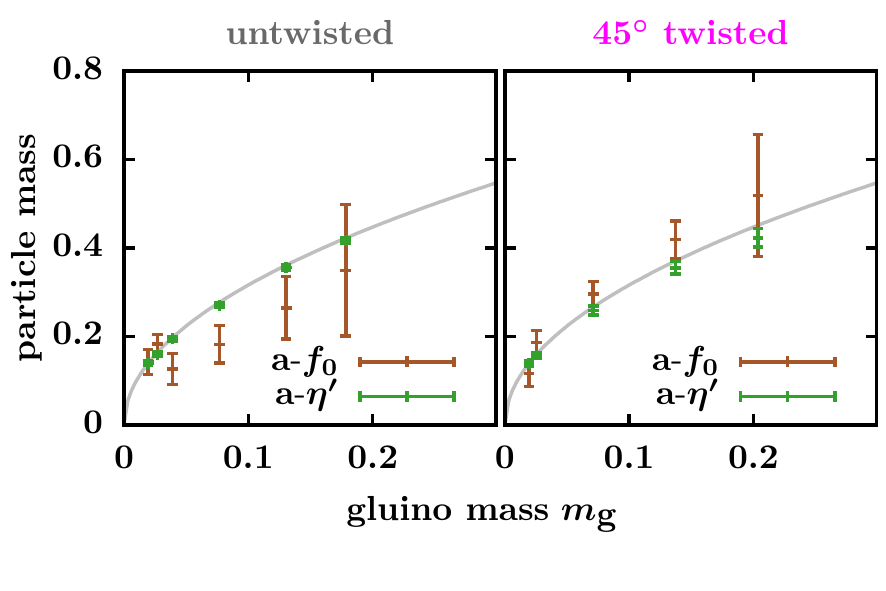}
		\caption{The masses of $\text{a-}\eta^\prime$ and $\text{a-}f_0$ for twist angles $\alpha\in\{0^\circ,45^\circ\}$ on the $8^3\times 16$ lattice.\\}
		\label{fig:EtaF8}
	\end{minipage}\hfill
	\begin{minipage}{0.47\textwidth}
		\flushright
		\includegraphics[width=\textwidth]{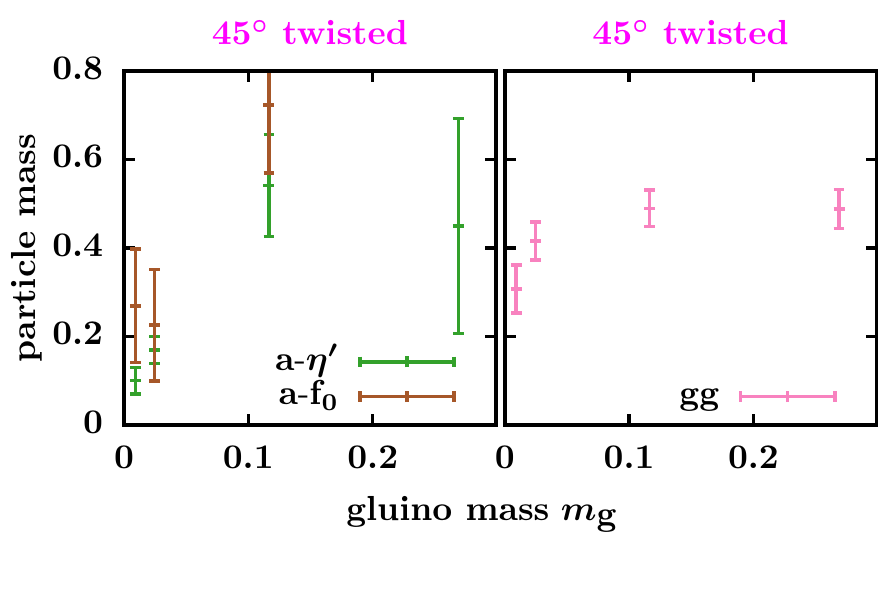}
		\caption{The masses of $\text{a-}\eta^\prime$ \& $\text{a-}f_0$ (left) and gluino-glue (right) for twist angle $\alpha=45^\circ$ on the $16^3\times 32$ lattice.}
		\label{fig:EtaFgg16}
	\end{minipage}
\end{figure}

The physical states of the Veneziano-Yankielowicz multiplet are numerically more demanding. The gluino-glue contains gauge fluctuations and the disconnected contributions of the mesonic states $\text{a-}\eta^\prime$ \& $\text{a-}f_0$ fluctuate quiet largely. Thus high statistics runs combined with gauge field (stout) and source (Jacobi) smearing are necessary. Figure~\ref{fig:EtaF8} shows the mesonic states of the supermultiplet calculated on the $8^3\times 16$ lattice. In comparison to Figure~\ref{fig:PiACuts} the uncertainty is bigger, but nevertheless the qualitative result is the same. Without a twist the $\text{a-}f_0$ is lighter than the $\text{a-}\eta^\prime$ and for $45^\circ$-twist both masses are degenerated within error bars.

To simulate a volume of approximately $(1\text{fm})^3$ we need at least a $16^3\times 32$ lattice. Our results are depicted in Figure~\ref{fig:EtaFgg16} and include the gluino-glue. The larger uncertainties are mainly a consequence of the (so far) smaller ensemble size.

\section{Summary}
Putting Super-Yang-Mills theory on the lattice breaks supersymmetry and states belonging to a supermultiplet have consequently different masses. The supersymmetry and its mass-degeneracy can be restored in the continuum limit. In our simulations with Wilson fermions we include a further parameter in the Dirac operator, which resembles the twisted mass formulation of lattice QCD. We found evidence that this approach eliminates at twist angle $\alpha=45^\circ$ the mass split of the chiral partners and thus improves the mass degeneracy of the supermultiplet at finite lattice spacing. As a consequence we expect a better extrapolation to the continuum limit and a faster restoration of supersymmetry. Our analytical considerations further support these numerical findings.

\subsection*{Acknowledgments}

We thank Philipp Scior for helpful comments, the Leibniz Supercomputing Center for funding the project pr48ji and the University Jena for providing the Ara cluster. M.S.\ received the student support of the conference.

\bibliographystyle{JHEP}
\bibliography{references}

\providecommand{\href}[2]{#2}\begingroup\raggedright\begin{thebibliography}{10}

\bibitem{Witten:1981nf}
E.~Witten, \emph{{Dynamical Breaking of Supersymmetry}},
  \href{https://doi.org/10.1016/0550-3213(81)90006-7}{\emph{Nucl. Phys.}
  {\bfseries B188} (1981) 513}.

\bibitem{Dimopoulos:1981zb}
S.~Dimopoulos and H.~Georgi, \emph{{Softly Broken Supersymmetry and SU(5)}},
  \href{https://doi.org/10.1016/0550-3213(81)90522-8}{\emph{Nucl. Phys.}
  {\bfseries B193} (1981) 150}.

\bibitem{Ellis:1983ew}
J.~R. Ellis, J.~S. Hagelin, D.~V. Nanopoulos, K.~A. Olive and M.~Srednicki,
  \emph{{Supersymmetric Relics from the Big Bang}},
  \href{https://doi.org/10.1016/0550-3213(84)90461-9}{\emph{Nucl. Phys.}
  {\bfseries B238} (1984) 453}.

\bibitem{Veneziano8206}
G.~Veneziano and S.~Yankielowicz, \emph{{An effective Lagrangian for the pure N
  = 1 supersymmetric Yang-Mills theory}},
  \href{https://doi.org/http://dx.doi.org/10.1016/0370-2693(82)90828-0}{\emph{Physics
  Letters B} {\bfseries 113} (1982) 231 }.

\bibitem{Shifman:1987ia}
M.~A. Shifman and A.~I. Vainshtein, \emph{{On Gluino Condensation in
  Supersymmetric Gauge Theories. SU(N) and O(N) Groups}},
  \href{https://doi.org/10.1016/0550-3213(88)90680-3}{\emph{Nucl. Phys.}
  {\bfseries B296} (1988) 445}.

\bibitem{Curci8612}
G.~Curci and G.~Veneziano, \emph{{Supersymmetry and the Lattice: A
  Reconciliation?}},
  \href{https://doi.org/10.1016/0550-3213(87)90660-2}{\emph{Nucl. Phys.}
  {\bfseries B292} (1987) 555}.

\bibitem{Ali:2019agk}
S.~Ali, G.~Bergner, H.~Gerber, I.~Montvay, G.~Münster, S.~Piemonte et~al.,
  \emph{{Numerical results for the lightest bound states in $\mathcal{N}=1$
  supersymmetric SU(3) Yang-Mills theory}},
  \href{https://doi.org/10.1103/PhysRevLett.122.221601}{\emph{Phys. Rev. Lett.}
  {\bfseries 122} (2019) 221601}
  [\href{https://arxiv.org/abs/1902.11127}{{\ttfamily 1902.11127}}].

\bibitem{Bergner:2019}
G.~Bergner, S.~Piemonte, G.~Münster, I.~Montvay, H.~Gerber, P.~Scior et~al.,
  \emph{{Continuum limit of SU(3) $\mathcal{N}=1$ supersymmetric Yang-Mills
  theory and supersymmetric gauge theories on the lattice}}, {\emph{{these
  proceedings}} {\bfseries \pos{PoS(LATTICE2019)175}} }.

\bibitem{Ali:2019gzj}
S.~Ali, G.~Bergner, H.~Gerber, S.~Kuberski, I.~Montvay, G.~Münster et~al.,
  \emph{{Variational analysis of low-lying states in supersymmetric Yang-Mills
  theory}}, \href{https://doi.org/10.1007/JHEP04(2019)150}{\emph{JHEP}
  {\bfseries 04} (2019) 150}
  [\href{https://arxiv.org/abs/1901.02416}{{\ttfamily 1901.02416}}].

\bibitem{Scior:2019}
P.~Scior, H.~Gerber, I.~Montvay, G.~Münster, S.~Piemonte, G.~Bergner et~al.,
  \emph{{Investigation of $\mathcal{N}=1$ supersymmetric Yang-Mills theory}},
  {\emph{{these proceedings}} {\bfseries \pos{PoS(LATTICE2019)038}} }.

\bibitem{Bergner:2019dim}
G.~Bergner, C.~López and S.~Piemonte, \emph{{Study of center and chiral
  symmetry realization in thermal $\mathcal{N}=1$ super Yang-Mills theory using
  the gradient flow}},
  \href{https://doi.org/10.1103/PhysRevD.100.074501}{\emph{Phys. Rev.}
  {\bfseries D100} (2019) 074501}
  [\href{https://arxiv.org/abs/1902.08469}{{\ttfamily 1902.08469}}].

\bibitem{Lopez:2019}
C.~Lopez, G.~Bergner and S.~Piemonte, \emph{{A study of thermal SU(3)
  supersymmetric Yang-Mills theory and near-conformal theories from the
  gradient flow}}, {\emph{{these proceedings}} {\bfseries
  \pos{PoS(LATTICE2019)235}} }.

\bibitem{Immirzi:1982hj}
G.~Immirzi and K.~Yoshida, \emph{{ Generalized Lattice Fermion Actions And
  Applications }},
  \href{https://doi.org/10.1016/0550-3213(82)90175-4}{\emph{Nucl. Phys.}
  {\bfseries B210} (1982) 499}.

\bibitem{bergner_low-dimensional_2008}
G.~Bergner, T.~Kästner, S.~Uhlmann and A.~Wipf, \emph{Low-dimensional
  supersymmetric lattice models},
  \href{https://arxiv.org/abs/0705.2212}{{\ttfamily 0705.2212}}.

\end{thebibliography}\endgroup


\end{document}